\documentclass[preprint,a4paper]{revtex4}
\usepackage{graphicx}
\begin{document}
\title{Electrical driving single barrier spin cell}
\author{B. H. Wu}
\author{Kang-Hun Ahn}
\affiliation{Department of Physics, Chungnam National University, Daejeon 305-764, Republic of Korea}
\begin{abstract}
We propose a spin cell based on
photon-assisted  tunneling through a conventional
semiconductor barrier. The Dresselhaus spin-orbit
interaction is included to break the spin rotation symmetry.
Due to the in-plane electric field induced asymmetric momentum distribution in one lead,
continuous flows of spin currents are driven through a barrier by a AC field.
The net charge current remains zero. The spin current via photon-assisted tunneling can be
readily adjusted via tuning the AC frequency or the in-plane electric field.
This device may function as an ideal spin cell to supply  spin currents in the spintronics circuit.
\end{abstract}
\maketitle
In recent years, an emerging technology named
spintronics has attracted much
attention (for review, see \cite{Science2941488,RMP76323}).
The motion of spin-polarized electrons produces
a spin-dependent particle current\cite{Science2941488,PRL90258301}.
To form a complete spintronics circuit, it is vital important to
have a spin cell which supplies continuous spin current through
the circuit.
Recently, such a spin cell was proposed based on the AC-driven
double-dot system in a spatially
\emph{nonuniform} external magnetic field\cite{PRL90258301}.
This spin cell can provide a spin current \emph{without} a charge current
by carefully adjusting the device parameters.
However, this spin cell met with some challenges such as the
nonuniform magnetic field.
An efficient non-magnetic spin cell is thus extremely desirable.
The spin-orbit interaction due to inversion asymmetry\cite{RMP76323}
 provides us the possibility to construct an
electrical driving spin cell based on the conventional semiconductors.

In this work, we propose an electrical driving single barrier spin cell,
where the $k^3$ Dresselhaus spin-orbit interaction
in a confined region is employed to break the spin rotation symmetry.
A continuous flow of spin
current is generated when the isotropic momentum distribution of electrons is broken by
an in-plane electric field, while the net charge current remains zero.
By applying an AC field,
the spin current can be greatly enhanced via photon-assisted
tunneling and  the spin injection can be readily
adjusted via tuning the AC frequencies or the in-plane electric field.
A tunable electrical driven spin cell  can be achieved.

The model device is illustrated in Fig. 1.
The device consists of a piece of zinc blende semiconductor potential barrier which is confined
by two ideal (free of charge and spin interaction) conducting leads.
The central region and the two leads are connected via tunneling.
Let the growth direction along $x\parallel [0,0,1]$ direction.
The coordinate axes $x,y$, and $z$ are set to be parallel to
the cubic crystallographic axes [0,0,1], [1,0,0], and [0,1,0].
The wave vector of electrons is $\mathbf{k}=(k_x,\mathbf{k}_\parallel)$,
where $\mathbf{k}_\parallel=
(k_y,k_z)$ is the in plane wave vector and $k_x$ is the wave vector along the growth direction.
We assume a perfect match at the interfaces of the device, where
$\mathbf{k}_\parallel$ is a good quantum number. The Hamiltonian can thus be
diagonalized in the basis $\mathbf{k}_\parallel$.
The central barrier we are interested is an open system where electrons can flow back and forth
between the barrier and the two leads. The dynamics of this open system can be well described
by an effective non-Hermitian Hamiltonian \cite{PRC6114}:
$\mathcal{H}_C=H_C-i\Sigma$,
where $H_C$ is the Hamiltonian of the central barrier
and the non-Hermitian term $-i\Sigma$ includes all the
boundary conditions such as the barrier-lead coupling.
By taking the assumption that the
incident electron's kinetic energy is much smaller than
the height of the barrier potential\cite{PRB67201304},
the central Hamiltonian $H_C$ of
the zinc blende semiconductor is then given in
the nearest neighbor tight-binding
approximation\cite{Datta} by
\begin{eqnarray}
  {H}_C&=&\sum_{n,\mathbf{k}_\parallel,\sigma}\tilde V_n(\mathbf{k}_\parallel)
  c^\dag_{n\mathbf{k}_\parallel\sigma}c_{n\mathbf{k}_\parallel\sigma}
  -t(c^\dag_{n\mathbf{k}_\parallel\sigma}c_{n+1,\mathbf{k}_\parallel\sigma}+H.C.)\\ \nonumber
  &&+\sum_{n,\mathbf{k}_\parallel,\sigma,\sigma'}(c_{n\mathbf{k}_\parallel\sigma}^\dag(-2\frac{\gamma}{a^2}
  (\hat\sigma_yk_y-\hat\sigma_zk_z)_{\sigma\sigma'})c_{n\mathbf{k}_\parallel\sigma'} \\ \nonumber
  &&+c^\dag_{n\mathbf{k}_\parallel\sigma}(\frac{\gamma}{a^2}(\hat\sigma_yk_y-\hat
  \sigma_zk_z)_{\sigma\sigma'})c_{n+1,\mathbf{k}_\parallel\sigma'}+H.C.),
\end{eqnarray}
where $\tilde V_n(\mathbf{k}_\parallel)$ contains all the on site
energies at $x=na$ ($n=1,2,...,N$), $a$ is the discrete lattice constant, $t=\hbar^2/2m^*a^2$
is the nearest neighbor hopping matrix element,
$\gamma$ is the Dresselhaus constant of the chosen material,
and $c^\dag_{n\mathbf{k}_\parallel\sigma}$
 ($c_{n\mathbf{k}_\parallel\sigma}$)
creates (annihilates) a spin $\sigma$ electron at site $n$ with a
given parallel wave vector $\mathbf{k}_\parallel$.
In the wide band approximation and symmetric coupling, the
non-Hermitian part of Hamiltonian has the form
$\Sigma=\sum_{\mathbf{k}_\parallel\sigma}\frac{\Gamma}{2}
c_{1\mathbf{k}_\parallel\sigma}^\dag c_{1\mathbf{k}_\parallel\sigma}
+\frac{\Gamma}{2}c^\dag_{N\mathbf{k}_\parallel\sigma}c_{N\mathbf{k}_\parallel\sigma}$, where
$\Gamma$ is the energy independent level width function.

For a high barrier height, the electron
transmission probability is greatly reduced by the nonresonant tunneling process.
We apply an AC field to achieve a tunable spin cell via photon-assisted
tunneling\cite{PR3951,PRL90210602}.
Suppose the junction is illuminated by a monochromatic
AC field of frequency $\omega$. The on-site energies of the
barrier are modified by the AC field. The total Hamiltonian is written
as $\mathcal{H}=\mathcal{H}_C+H_{AC}$, where $H_{AC}=-\sum_{n\sigma}
A_m\cos(\omega t) c^\dag_{n\mathbf{k}_\parallel\sigma}
c_{n\mathbf{k}_\parallel\sigma}$.

In this work, we generalize the Floquet-Green
formalism\cite{PRL90210602}  to the spin-involved transport problems.
For a given in-plane wave vector $\mathbf{k}_\parallel$, solution of
the time-dependent Schr\"odinger equation in the Floquet ansatz takes
the form $|\psi_m(t)\rangle=\exp[-i\epsilon_mt/\hbar]|\varphi_m(t)\rangle$,
where $\epsilon_m$ is the quasienergy. The Floquet states $|\varphi_m(t)\rangle$
can be expanded in Fourier expansion form as $|\varphi_m(t)\rangle=
\sum_p|\varphi_{m,p}\rangle\exp[-ip\omega t], p\in Z$. Both the
Floquet states and the quasienergies are determined by the eigenvalue
equation $\mathcal{F}|\varphi_m(t)\rangle=\epsilon_m|\varphi_m(t)\rangle$
with a non-Hermitian operator $\mathcal{F}=\mathcal{H}-i\hbar\frac{\partial}
{\partial t}$ \cite{footnote1}.
Since the eigenvalues are generally complex,
we have to solve the adjoint
equation $\mathcal{F}^*|\phi_m(t)\rangle=\epsilon^*_m|\phi_m(t)\rangle$ to
form a complete biorthogonal basis.
The propagation of an electron from
site $n'$ with spin $\sigma'$ and energy $\epsilon$ to the state at
site $n$ and spin $\sigma$ under the absorption ($p>0$) or emission ($p<0$)
$|p|$ photons is described by the retarded Green function
\begin{equation}
G^{p}_{n\sigma,n'\sigma'}(\epsilon,\mathbf{k}_\parallel)=\sum_{m,p'}
\frac{\langle n\sigma|\varphi_{m,p+p'}\rangle
\langle\phi_{m,p'}|n'\sigma'\rangle}{\epsilon-\epsilon_m-p'\hbar\omega}.
\end{equation}

After obtaining the Green functions, spin-involved transport properties
can be described by the 2$\times$2 $p$-photon assisted transmission matrices
$\mathbf{T}^p_{LR}$ and $\mathbf{T}^p_{RL}$,
where the matrix element $T^p_{\alpha\beta,\sigma\sigma'}$ denotes the
time-averaged $p$-photon assisted transmission
probability for electrons from the $\beta$ lead to
the $\alpha$ lead with the initial and final spin state
being $\sigma'$ and $\sigma$ respectively. These transmission matrices
can be easily obtained by generalizing the result in Ref. \cite{PRL90210602}
to the present spin dependent transport problem.
The diagonal and non-diagonal elements  characterize,
respectively, the spin-conserved and the spin-flip transport probabilities.
Our model has a symmetric design and a standing wave AC potential.
The two transmission probabilities are equal since the electron
has no way to distinguish the left lead or the right lead. We therefore
omit the subscript of the transmission probability as
$\mathbf{T}^p(\epsilon,\mathbf{k}_\parallel)$ for simplicity.

We apply an external electric field $\mathbf{F}=(F\cos\theta,
F\sin\theta)$ in the $y$-$z$ plane in the left lead to create
an anisotropic lateral momentum distribution.
We assume the central barrier is much higher than the
potential caused by the electric field. In the
weak electric field limit, the tunneling correction to
the distribution function in the lead can be neglected.
The difference between the left and the right distribution function in
the relaxation time approximation is given by\cite{Galp,JAP87387}
\begin{equation}\label{f1}
  f_L- f_R\approx e\tau_{tr}\cdot
  \frac{\partial f_0}{\partial \epsilon}
 (\mathbf{v_\parallel}\cdot\mathbf{F}),
\end{equation}
where  $\tau_{tr}$ is the
transport relaxation time for electrons and $\mathbf{v}_\parallel=\frac{\hbar
\mathbf{k}_\parallel}{m^*}$ is the
in plane electron velocity. At low temperature ($T\rightarrow 0$), the derivative
$-\frac{\partial f_0}{\partial\epsilon}$ can  approximately be replaced by a
simple $\delta-$function. The spin current can be found from the time evolution
of the spin-resolved occupation number operator and
the spin current density $j_\sigma$ (in unit of A/m$^{2}$) is given by
\begin{equation}\label{Jspin}
j_\sigma=\frac{e^2\tau_{tr}}{8\pi^3m^*}\int_{0}^{k_F}k_\parallel
dk_\parallel\int^{2\pi}_0d\eta(\mathbf{k}_\parallel\cdot\mathbf{F})
\sum_{p\sigma'}{T}^p_{\sigma\sigma'}
\end{equation}
where the in plane wave vector $\mathbf{k}_\parallel=(k_\parallel\cos\eta
,k_\parallel\sin\eta)$ ($\eta$ is the polar angle)  and the Fermi wave vector $k_F$
are determined by the carrier density in the leads.

We numerically investigated the spin cell model with realistic parameters.
We choose the material of central
junction to be GaSb which has a larger Dresselhaus constant $\gamma$ than other
III-V materials\cite{PRB67201304}. The material parameters
are given by$\gamma=187$ eV$\cdot$\AA$^3$ and $m^*=0.041m_0$, where $m_0$ is
the electron mass.
The length of the central junction in our device is $L=5a=100$\AA.
The AC potential parameter is fixed at $A_m=0.1t$. The Fermi
energy of leads and the barrier on site energy are fixed at 0 and $4t$, respectively.
The carrier density at two leads is set to be
$2\times 10^{17}$ cm$^{-3}$ and the Fermi wave vector
is estimated to be $1.8\times 10^8$ m$^{-1}$. A relaxation time of 1 ps is used
in our numerical calculations.

Fig. 2 depicts the surface spin current density as a function of the driving
field frequency for a given external in-plane electric field.
The in-plane electric field is $F=1.6\times 10^3$ V/m with the polar angle $\theta=\pi/3$.
The left lead electron distribution in the $(k_y,k_z)$ space is therefore no longer isotropic.
From the symmetry properties of our Hamiltonian, we have $T^\sigma(\epsilon,\mathbf{k}_\parallel)
=T^{-\sigma}(\epsilon,-\mathbf{k}_\parallel)$. Therefore, the spin current densities
satisfy the relation: $j_\sigma=-j_{-\sigma}$.
The two spin currents have the same magnitude but opposite signs.
Therefore, the net charge current is always zero.
When the frequency of the AC field matches the energy
difference between the Fermi energy and the  resonant level of the central junction,
the spin current reaches its maximum value due to photon assisted tunneling.
When the AC frequency is far off the resonant frequencies, the AC pumped
spin currents decrease.
Due to the spin-orbit interaction, the degeneracy of the lowest resonant level is lifted.
One can find two photon-assisted tunneling resonant spin current peaks in Fig. 2.
By scanning the frequency across the
two resonant frequencies, the spin currents will reverse their direction.
Therefore, we can readily control the direction and the
magnitude of the spin current of the proposed spin current source
by tuning the AC field frequency.
If we decouple our device with the external circuit, spin accumulation occurs at the
edges of the device due to the balance of spin accumulation and spin relaxation.
This spin-accumulation may be experimentally detected by optical techniques
based on Kerr/Faraday rotation with submicron spatial resolution\cite{Science3061910}.

The output spin current can also be tuned by the in-plane electric field.
In Eq. (\ref{Jspin}), the spin current density has a trivial
proportion relation with the amplitude of the electric field.
However, the polar angle of the in-plane electric field provides
us another effective means to control the spin currents.
In Fig. 3, we show the numerical calculated surface spin current density as a function of the
polar angle $\theta$ of the in-plane electric field.
One can see that the spin currents are modulated by
$\theta$ in a sinusoidal way.
This sinusoidal modulation is caused by
the general feature and symmetry of the field induced nonequilibrium distribution
and the spin-orbit interaction. We note the spin quantization direction is along $z$ axis.
From Eq. (\ref{Jspin}), one can find that $j_\sigma$ is modulated by $\cos\theta$.
For $\theta=n\pi,\ n\in Z$, i.e.
the electric field is vertical to the $z$ direction (in $y$ direction), the spin
currents are suppressed to zero.
However, when the in-plane electric field deviates from $y$ axis,
a non-vanishing spin current will be generated.
Especially when the electric field is along the $z$ axis ($\theta=\frac{2n+1}{2}\pi$),
this spin current reaches its maximum.

In conclusion, we have studied an AC driven spin cell
by generalizing the Floquet-Green function method.
The $k^3$ Dresselhaus spin-orbit interaction due
to bulk inversion asymmetry contributes to the spin-dependent effective mass correction
and leads to the splitting of the resonant level.
Driving the lead electron distribution out of equilibrium by an in-plane electric field,
a continuous flow of
spin current can be generated by the AC field while the net charge current
remains zero. The direction and the magnitude of the generated spin current can be readily
tuned by changing the AC frequency or the in-plane electric field direction.
Besides all the basic characteristics of spin cell presented in Ref. \cite{PRL90258301},
our spin cell functions without external magnetic field or any ferromagnets.
It can be operated in a fully electrical way.
By integrating our proposed spin cell with spin information processors,
one can form a complete spintronics circuit based {only} on conventional semiconductors.

This work was supported by Korea Research Foundation, Grant
No.(KRF-2003-070-C00020).

\newpage
FIGURE CAPTIONS:\\

FIG. 1: Sketch of the proposed spin current source. The material
growth direction is along $x$ axis. The left lead electrons are driven
by an in-plane ($y$-$z$ plane) electric field and the central
junction is illuminated by an AC field.

FIG. 2: The surface spin current densities as a function of the driven
frequency (in unit of the hopping matrix term $t=\hbar^2/2m^*a^2$)  for a given external in-plane electric field.

FIG. 3: Numerical calculated surface spin current densities versus the
polar angle $\theta$ of the in-plane electric field at a fixed frequency
$\hbar\omega=2.26 t$.

\newpage
\begin{figure}
\includegraphics{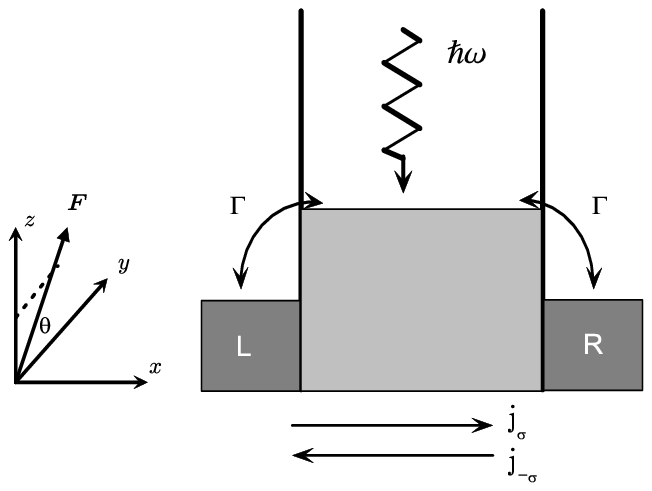}
\caption{by B.H. Wu, and Kang-Hun Ahn}
\end{figure}

\newpage
\begin{figure}
\includegraphics{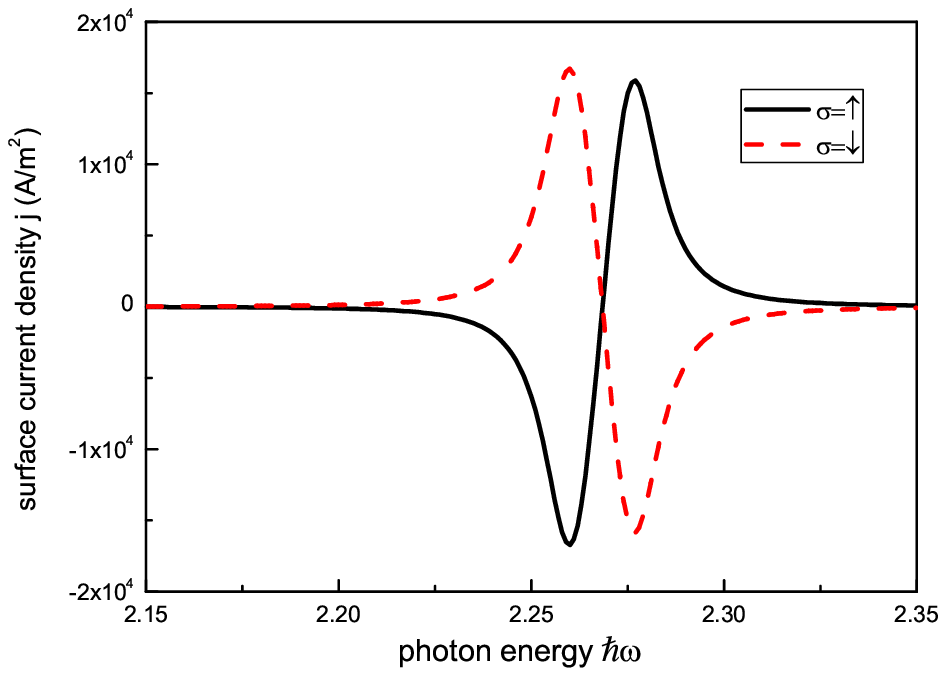}
\caption{by B.H. Wu, and Kang-Hun Ahn}
\end{figure}

\newpage
\begin{figure}
\includegraphics{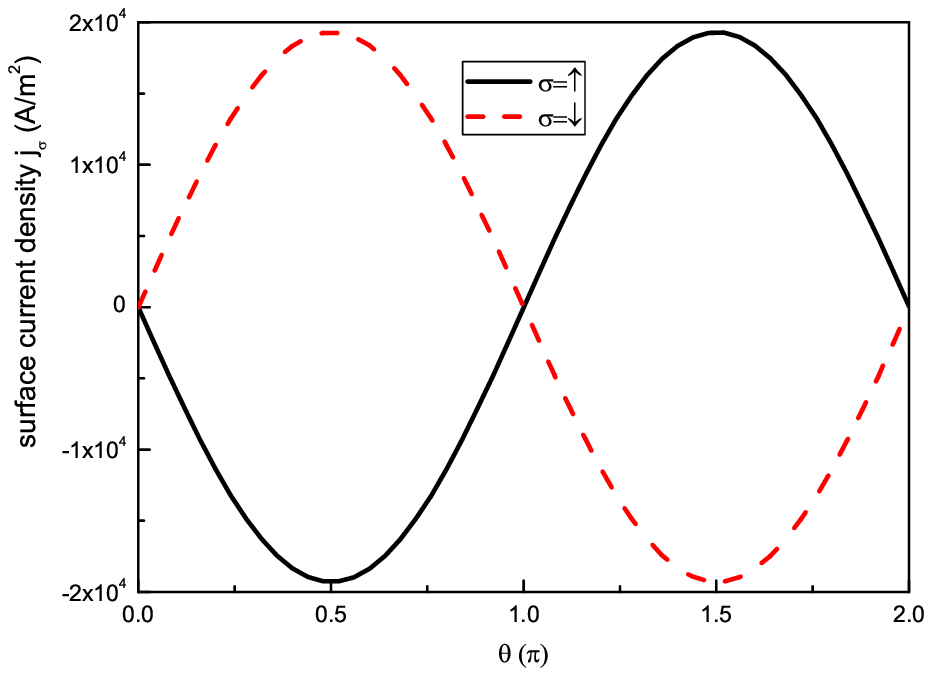}
\caption{by B.H. Wu, and Kang-Hun Ahn}
\end{figure}

\begin{thebibliography}{9}
\bibitem{footnote1} The completeness of the eigenvectors of non-Hermitian Hamiltonian is determined by the multiplicity
of the physical and algebraic degeneracy. In the case of spin, it is always possible to lift the spin degeneracy
by a small $\eta\rightarrow 0$ variable, e.g. a small Zeeman splitting.
\end{thebibliography}
\end{document}